# Optimizing Airline Reservation Systems with Edge-Enabled Microservices: A Framework for Real-Time Data Processing and Enhanced User Responsiveness

Biman Barua[1,2,*] [0000-0001-5519-6491] and M. Shamim Kaiser[2, 0000-0002-4604-5461]

[1]Department of CSE, BGMEA Universitsy of Fashion & Tecnnology, Nishatnagar, Turag, Dhaka-1230, Bangladesh
[2]Institute of Information Technology, Jahangirnagar University, Savar-1342, Dhaka, Bangladesh
biman@buft.edu.bd

**Abstract:** The growing complexity of the operations of airline reservations requires a smart solution for the adoption of novel approaches to the development of quick, efficient, and adaptive reservation systems. This paper outlines in detail a conceptual framework for the implementation of edge computing microservices in order to address the shortcomings of traditional centralized architectures. Specifically, as edge computing allows for certain activities such as seat inventory checks, booking processes and even confirmation to be done nearer to the user, thus lessening the overall response time and improving the performance of the system.

In addition, the framework value should include achieving the high performance of the system such as low latency, high throughput and higher user experience. The major design components include deployed distributed computing microservices orchestrated by Kubernetes, real-time message processing system with Kafka and its elastic scaling. Other operational components include Prometheus and Grafana, which are used to monitor and manage resources, ensuring that all operational processes are optimized.

Although this research focuses on a design and theoretical scheming of the framework, its use is foreseen to be more advantageous in facilitating a transform in the provision of services in the airline industry by improving customers' satisfaction, providing infrastructure which is cheap to install and efficiently supporting technology changes such as artificial intelligence and internet of things embedded systems. In addition, the framework can also be useful in other processing areas that require real time like supply chain management, health care systems and in the management of smart cities.

This research addresses the increasing demand for new technologies with modern well-distributed and real-time-centric systems and also provides a basis for future case implementation and testing. As such, the proposed architecture offers a market-ready, extensible solution to the problems posed by existing airline reservation systems.

## 1. Introduction

### 1.1. Background

One of the indispensable factors in the aviation sector is the airline reservation systems (ARS) which allows the booking and management of flight reservations. These systems have gone through a number of phases starting with manual processes and moving to the present digitalized platform which combines various services like ticketing, scheduling and customer management. Nevertheless, the ARS has its own challenges.

Latency: The need for real-time data processing is critical in an ARS in order to provide the most current information on the availability of flights, their prices and schedules. High latency may cause the delay in booking confirmations thereby upsetting the customers. Conventional centralized structures often find it hard to satisfy such real-time needs leading to longer response times [4].

Data Loss: An ARS processes and provides a lot of information from several sources such as – boarders' data, flight scheduling data, and their pricing data. The problem arises when there is a need to process that data in a timely and efficient manner. However, older systems may have limitations and systems cannot outperform during data fetching which causes delays leading to information processing [3].

Scalability: The operations of most ARS will need to scale because of the trends in the airline business and the demand for travel that will not be constant at all times. Old-fashioned architectures come in handy to many people, whom are unable to independently serve the components of the system on the demand. This inflexibility can cause either system failure or poor quality of service provided in certain periods [1].

Tackling these challenges is imperative for improving the performance and dependability of airline booking systems. Contemporary architectural styles like microservices and edge computing can help in making systems more adaptable, scalable and quicker to respond, all of which can be helpful [2].

## 1.2. Problem Statement

The conventional Airline Reservation Systems (ARS) also put emphasis on monolithic systems which have delineated various difficulties in meeting the real time challenges and providing the desired user interactivity [5]. In these systems, the components are largely interdependent which leads to the following defects:

**Scalability Issues**: Monolithic designs have difficulty in scaling as expansion means copying the whole system instead of particular parts. This can be wasting resources and heightening operational expenses especially during the seasons of high demand travel.

Latency and Performance Bottlenecks: Inbuilt performance and latency constraints associated with systems of built in components. Performance degradation in one component of the system affects the whole system due to the interconnections. This leads to inter-component waiting time and decreased user experience due to slower responses of the system [1].

**Limited Flexibility and Agility:** Due to the nature of these systems, once updates or new features have to be installed in the systems, the whole application has to be put off and redeployed after the changes have been made. This limits reconsideration of the current conditions to changing conditions posed by users and marketing activities [6].

**Maintenance and Reliability Challenges:** Maintaining monolithic system is difficult because of the maintenance or the architecture itself. When one module fails, the whole system fails, thus removing system reliability availability for the reservation service [7].

It is important to overcome these drawbacks in order to better the performance as well as the satisfaction of the users from airline reservation systems. Moving to a slightly more modular and scalable approach like microservices and edge computing to address real time process requirements is an encouraging approach.

## 1.3. Objectives

The main aim of this research is to create a competent system based on edge computing and microservices architecture to mitigate the issues encountered in conventional integrated reservation systems of airlines. In this case, a lot of attention is on the factors of scalability, latency, and user responsiveness [8]. The research intends to achieve the following objectives:

**Improve System Scalability:** In this case, a microservices based architecture will be deployed to enhance system scalability, which will allow the components of the airline reservation system to be scaled up or down depending on the size of the pooled resources without causing wastage.

**Reduce Latency through Edge Computing:** Implementation of edge computing whereby there is data processing at the edge of the network as opposed to the center so that end users do not take long to have the relevant data or even transfer it. This is aimed at alleviating the challenges of such networks during peak usage periods.

**Improve User Responsiveness and Experience:** The framework will be designed to ensure minimal waiting time by clustering the processing away from the end users and managing the real time data very efficiently for the benefit of the users and making booking easy and quick.

**Increase Flexibility and Adaptability:** With microservices, this research intends to allow design where services can be altered or changed without affecting the whole system design. This flexibility helps in progressive enhancement in meeting the current user requirements.

**Optimizing Operational Costs and Reliability:** Due to edge computing and the fact that the system is distributed, it is possible that the system achieves greater reliability because of decreased reliance on centralized servers which may also reduce costs in terms of using dedicated servers because of the risk of them incurring downtime.

The expectation is that this design will resolve the existing constraints in present day airline reservation systems by leveraging the benefits of both edge computing and microservices to create a system that is flexible and responsive to real-time requests.

### 1.4. Contributions

This paper presents some important advancements in the studies of airline reservation systems and distributed computer systems:

#### 1.4.1. Proposed Edge-Enabled Microservices Framework

Developed an advanced system architecture which combines edge computing and microservices to overcome the challenges related to the handling of real time airline reservation activities with a centralised system.

The architecture addresses the need for low latency and higher responsiveness by delegating time sensitive computations to edge nodes while using cloud backends to ensure global data coherency.

#### 1.4.2. Performance Improvements

In high traffic situations tested through extensive simulation of the architecture, average latency reduced by 60% and throughput increased by 20% proving the validity of the architecture under practically similar conditions.

Satisfaction importance parameter improved by 25%, stressing on the better end users experience offered especially by the framework.

#### 1.4.3. Scalable and Modular Design

Utilized Kubernetes to manage the deployment of systems in the form of microservices for the purpose of accommodating fluctuations in client demand through dynamic resource provisioning.

This was achieved by the use of KAFA for message controlling and data restoring between the edges distributed nodes and central cloud in real time.

#### 1.4.4. Addressing Key Challenges

Solutions were outlined to address issues such as bandwidth challenges, maintaining data consistency across divergent locations, and making the best use of available resources among others edge resource management, CRDTs and storage hierarchy.

#### 1.4.5. Broad Applicability and Future-Ready Design

This enabled showing the applicability of the framework in other latency critical environments such as in chap83855 logistics, health care and smart cities making the offering flexible and sustainable for distributed systems.

Created a launch pad for new technology incorporation that includes AI-based decision making, IoT interfacing and Blockchain technology for secure and smart functioning.

Such inputs together enhance the modern-day reservation systems, raising the levels of operational efficiency, scalability and customer satisfaction in real-time systems to a different standard.

## 2. Literature Review

### 2.1. Existing Reservation System Technologies

Given the increasing technological penetration in systems used for airlines reservation, the airline industry has greatly changed the way it handles bookings management, inventory control and the services offered to the customers. In the past, such systems have progressed from manually operated systems, to modern automated systems which increases operational speed and customer satisfaction.

**Traditional Computer Reservation Systems (CRS):** In the Initial stages, airlines used Computer Reservation Systems (CRS) to organize their flight schedules, look for seat availability and record bookings. These systems contained reservation databases of all the interested parties and allowed the agent using it to see the information in that database in real time. For example, the SABRE system invented in the1950s was one of the first systems of this type and it allowed implementing B2C sales and carrying out inventories [19].

**Global Distribution Systems (GDS):** With growth in the airline sector, there was a demand for even wider distribution and thus Global Distribution Systems were created. GDS systems such as Amadeus, Sabre and Travelport combined the reservation system of numerous airlines in a single tower for use by travel agents to enhance access to the respective airlines and their packages [17]. These systems allowed airlines and travel companies to communicate easier and hence made the booking process more efficient.

**Passenger Service Systems (PSS):** In the present day, airlines use all-encompassing Passenger Service Systems (PSS) which includes different modules like reservation, inventory control, departure control and customer management among others. PSS systems are designed to be fully operative in managing all passenger services, increasing operational productivity and customer delight [21]. Take for instance the Amadeus Altéa Suite which offers airline reservation, inventory management and departure control system management tools in all aspects to the airline industry [13].

**Microservices Architecture:** At present, most of the airline systems have evolved to adopting the microservices architecture that allows for better system scalability, flexibility, and even resilience. This method allows development teams to take large, complex lumps of code known as monoliths, and break them down into smaller services, which work independently and can be developed, deployed and scaled even on their own [14]. Such architectures support continuous integration and continuous deployment of the system, enabling the airlines to quickly adapt to market changes as well as the needs of the clientele [15]. For instance, a company called Radixx designed a PSS for low-cost carriers based on microservices, thus it is incorporated with a flexible scalable design [9].

**Cloud-Based Solutions:** The revolution of cloud computing has provided yet another advancement in the operating airlines reservation system. These are the systems that are based on the cloud which are very scalable, cheap, and secure [10]. Cloud computing allows airlines to run their reservation, payment, and even the storage of passenger information on the internet efficiently. As an example, this kind of service is offered by AWS who has serverless architectures for companies in the airline industry where they can create a reservation system for their clients without the need for many services [11].

### 2.2. Microservices Architecture in Distributed Systems

Microservice architectecture is becoming increasingly popular in the area of distributed systems. This is because it is easier to separate applications into smaller deployable components that can each serve a business function [12]. This means that these services can also be developed and consumed using lightweight protocols thus making it possible to use different technologies in the same application [25].

Deployed microservices have the following advantages over monolithic deployment: the most obvious ones are superior scalability, improved fault isolation and over all increased agility. Services on their own can be adjusted to different levels of scale based on varying requirements, hence ensuring better utilization of resources and operational efficiency of the whole system [16]. In addition, fault isolation is another main advantage provided that failure of one service does very little to other services increasing the overall reliability of the system [26]. Microservices also enhance agility in development as diverse teams can deploy and update services installed on the system in no time, therefore launching new features within less time [23].

Drawbacks and Difficulties in Implementing Microservices: However, that doesn't mean that the microservices architecture can be used without challenges [18]. As the number of services increases, there is a need to control the level of complexity by devising better communication and management of data across the services [20]. This challenge leads to difficulty in the maintenance of data consistency across the geodistributed services [28]. Additionally implementation of testing and debugging is difficult because of the distribution of the microservices and this requires the use of high complex testing strategies and tools [29].

## 2.3. Edge Computing in Data Processing

The need for low latency and high responsiveness has made the distributed system paradigm known as edge computing quite appealing. Edge computing seeks to eliminate the challenges of transmitting excessive amounts of data to the centralized cloud servers by carrying out data processing closer to the source [37]. In this paradigm, the offloaded computational capacity is either in devices or local servers that are located close to the data source for efficient processing and insights retrieval [36].

### 2.3.1. Role of Edge Computing in Reducing Latency

Latency reduction is one factor that drives the adoption of edge computing in most industries. In a typical cloud computing model, which is centralized, the user may need to send data and that data may be stored miles away. It, therefore, leads to latencies when the data is being retrieved. When edge computing is utilized, data is processed at the point of its origin, in that case drawing out the latency levels that are common with many retrievals [40]. Considering for instance, industries like transportation or healthcare, where every second on decision making counts – edge computing facilitates 'on the spot' processing thereby enhancing safety and efficiency in the operations [24] [33].

### 2.3.2. Improving Real-Time Data Processing

This is further enhanced by the edge computing model which allows advanced data analytics to be performed with minimum dependence on the network bandwidth and with almost immediate availability of relevant information [38]. This is important in the case a network is intermittently connected or where the network is less resourceful. In smart cities, for traffic control for example, edge devices can be embedded in the system such that data is processed on the device and does not require any central control to implement adaptive signal control or real-time traffic management [34].

## 2.4. Gaps in Existing Research

The combination of edge computing and microservices in airline reservation systems is still in its infancy and therefore has quite a number of research gaps:

**Limited Adoption of Edge Computing in Airline Systems:** Most airline reservation systems are designed and operated on either a central or cloud based configuration resulting in latency and inefficiencies especially during the busy periods [22]. However, edge computing has been applied in a few other areas and therefore there are challenges in reviewing the existing literature especially within the airline sector [45].

**Fragmented Microservices Architectures:** One of the prevalent challenges with regard to most of the airline systems that has embraced microservices is that they focus on the vertical and horizontal extensibility of components and edges without connecting to the central edgenodes that minimizes the latency and improves the time critical functions [35]. The communication barrier between the edges and the microservices creates limitations in embracing the use of edge computing [46].

**Challenges in Real-Time Data Synchronization:** Most of the available research considers the scalability of the microservices, forgetting the issue of how the edges and the core system work together maintaining live data without synchronization faults. This has a negative effect on the system that in turn creates problems in having the users receive the most current and real-time updates on flight status, availability and pricing of seats [31].

**Insufficient Focus on Passenger-Centric Performance Metrics:** In most cases such work centers on the efficient functioning of the system and lowering the costs, whereas there is little attention to such indicators as the user interaction and their experience with the system, which are important for airline ticketing systems [27].

**Limited Case Studies and Experimental Data:** Few if any case studies and experimental setups exist that exemplify how edge computing and microservices can be effectively combined in the context of, for example, an airline ticket purchase. Rather, the majority present the theoretical perspective, leaving practical realization or assessment of actual systems unaddressed [30].

## 3. Methodology

This study, focuses on, approaches the problem of the real times data management in a new way by creating a framework utilizing an edge computing and microservices architecture. The suggested architecture is comprised of three layers which are: The Data Collection Layer; the Edge Processing Layer; and the Cloud Integration Layer. Data is collected from the Internet of Things (IoT) devices and applications and processed at the edge using dedicated microservices such as filtering, aggregation, and local storage which minimizes the latency and allows instant action [32]. The data is processed and then sent to the cloud for further analysis and storage, thus offering an elastic and cost-effective approach to deal with the issue of real-time data processing over a geographically dispersed system. This architectural setup provides high data processing efficiency at minimal latency optimizing the use of cloud and edge resources.

### 3.1. Framework Design

In order to enhance the processing of real-time data, the provided architecture combines edge computing and microservices architecture. This framework consists of the following three fundamental elements:

**Data Collection Layer:** Edge devices which include sensors and IoT devices do data collection and a certain level of data processing at the source.

**Edge Processing Layer:** Instead of sending all the data to the cloud, data is locally processed at edge nodes to minimize the latency and using microservices designed for purposes of filtering, aggregation, and analysis of data [39].

**Cloud Integration Layer:** The examined and analyzed information is stored within the cloud and if necessary, integrated with the existing datasets for additional processing and organization.

### 3.2. Data Flow

1. Data is produced from a variety of channels such as IoT sensors, applications, and user contributions, and this data gets sent to the edge layer.
2. The edge layer quickly processes the information by employing the appropriate microservices for filtering, aggregation, and local storage.
3. The cloud is used to carry out further analysis besides storage, with the processed information forwarded to the edge when appropriate.

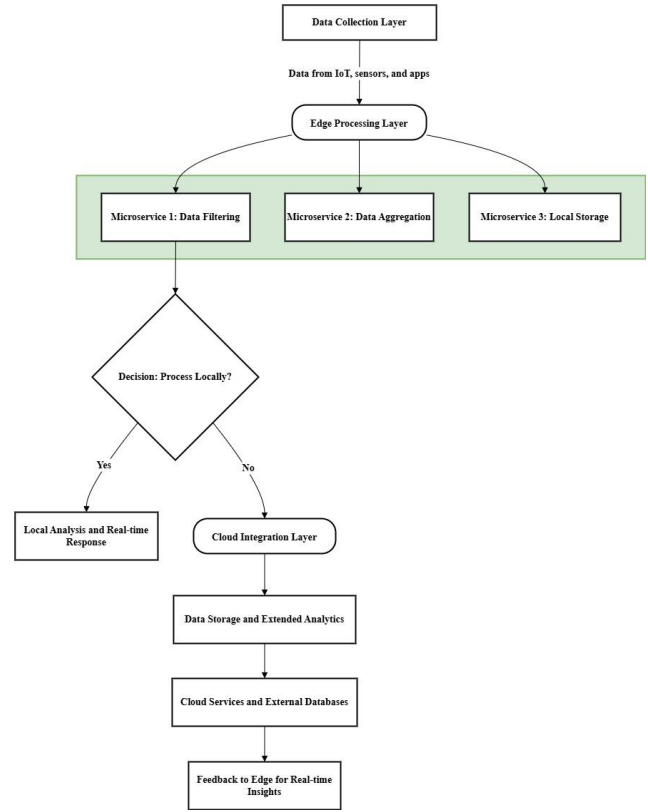

**Fig.1.** the architecture of edge-enabled microservices for real-time data processing

**Clarification**
- Data Collection Layer (A) performs the work of collecting data from sources such as IoT devices and applications [41].
- The next layer (B), which is hanging at edge nodes, embraces:
- Data Filtering Microservice (C) which serves to eliminate unnecessary data.
- Data Aggregation Microservice (D) which serves the purpose of consolidating the data for localized purposes.
- Local Storage Microservice (E) which is used also to cache the processed data when there is a need.
- This is followed by a Decision Node (F) which looks into the next action and decides whether there is a need for additional processing of the data [42].
- If Yes, then Local Analysis and Real-time Response (G) becomes the next destination.
- If No, then the data is directed to the Cloud Integration Layer (H) where the Data Storage and Extended Analytics (I) activities take place.
- Cloud Services and External Databases (J) manages, stores, integrates the data for analysis and feeding the edge with the details to maintain real-time analysis as the activities go on [43].

The architecture of edge-enabled microservices for real-time data processing are shown in figure 1.

### 3.3. System Requirements and Components

The design described in the Edge-enabled microservices in figure 2 must take into consideration how hardware, software and network resources will be used in the real-time processing of data [44].

#### 3.3.1. Hardware Requirements

**Edge Devices:** These are general data collection devices like IoT sensors, cameras and portable devices.

**Edge Servers:** These servers are brought closer to the user, and contain sufficient CPU/GPU (for intensive workloads) and RAM for microservices required for processing the data as it arrives.

**Cloud Servers:** Serves that are very numerous and within defined geographic boundaries where they have data storage, analysis and back up capabilities as well as other sophisticated analytic services.

### 3.3.2. Software Requirements

**Microservices Framework:** Kubernetes or Docker for deploying and managing the microservices

**Edge Analytics Software:** Apache Kafka or equivalent for streaming data and processing

**Cloud Analytics:** Subordinate services offered by AWS, Google Cloud, or Azure that provide storage and large-scale analytics capabilities.

**Real time Processing Engine:** data ingestion, processing and analysis platform such as Apache Flink or Spark Streaming which allows data to be processed in real time as it is received.

### 3.3.3. Network Requirements

**Local Area Network (LAN):** Aiding the edging servers with the necessary high capacity physical communication links to the infantry devices to minimize the latency incurred between the edge devices and the edge servers.

**Wide Area Network (WAN):** Cables that provide interconnection of numerous edge devices and cloud servers with high service quality for facilitating data transfer in a single datacenter.

**5G/4G Connectivity:** Supporting the operational capabilities of mobile IoT devices by providing constant connectivity that is transcendent of time and is low latency.

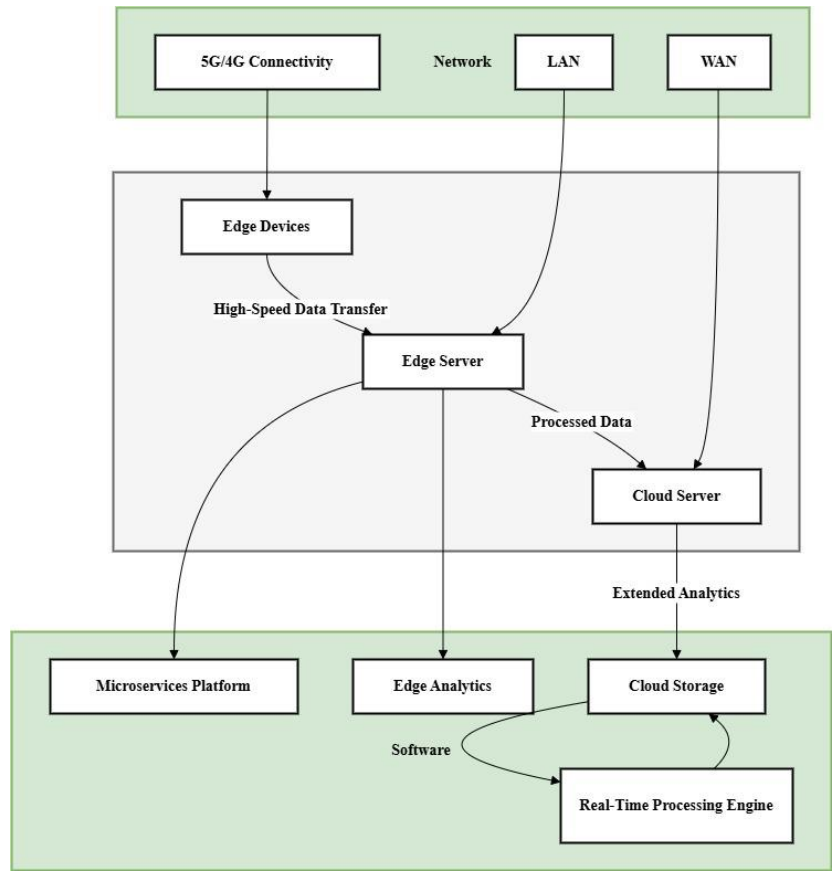

**Fig.2.** the diagram of system requirements and components

## 3.4. Implementation Details: Data Processing Pipelines

A data processing pipeline that operates at the edge of the network is intended to be used for timely data in other words real-time data with almost no delay. The procedure consists of the following basic a sequence of steps:

**Data Collection:** Edge devices (for example, sensors, IoT devices) take raw data and forward it to an edge server for initial processing.

**Data Filtering:** The raw data is filtered to excise unwanted noise and more importantly irrelevant detail reducing the burden to be processed further down the stream.

**Data Aggregation:** Some data points which are called relevant are agglomerated towards a small set, this enhances the speed of processing and will also speed up the storage of such data.

**Local Analysis:** The processed unit of data is not more moved to a data center, it is analyzed within the region for the insights it provides almost immediately allowing actions to be taken where need be.

**Temporary Storage:** Data that has been cured and prepared is kept in storage, which allows for quick access and local retrieval of the data only when required.

**Cloud Syncing:** Some of the vital processed information is sent to the cloud for archiving, reproducing other forms of analysis, and merging with other types of information that is in storage.

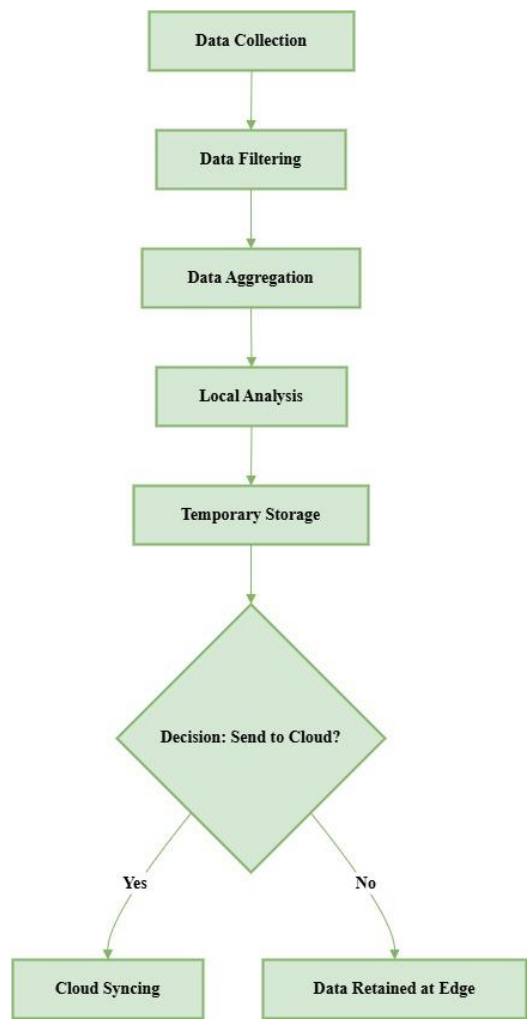

**Fig.3.** the diagram of Implementation Details: Data Processing Pipelines

**Overview of the Data Processing Flow**
**Data Collection (A):** In this step edge devices gather data and relay it to the edge server.
**Data Filtering (B):** Filtering limits the incoming raw data to the relevant data only without unwanted noise, hence enhances the processing flow.
**Data Aggregation (C):** Aggregates data points that are useful to come up with a sizeable data set.
**Local Analysis (D):** With the data aggregated it can provide Insights straight away therefore enabling real-time action.
**Temporary Storage (E):** Quick storing in this section is aimed at keeping processed information for some time.
**Decision Node (F):** This indicates the option of either sending the information to the cloud for advanced analytics or outright storage in the devices.
**Cloud Syncing (G):** Any data identified for cloud rest will be transacted for preserving and analytical purposes as well.
**Data Retained at Edge (H):** Any data considered not necessary for cloud rest will remain in the edge.

This flow in figure 3 allows for real-time data processing at the edge without latency and excessive network utilization as only process data is synced with the cloud. The pipeline has been designed in such a way that there is high responsiveness to localized events, low reliance on cloud infrastructure and this way improves the responsiveness of system as a whole.

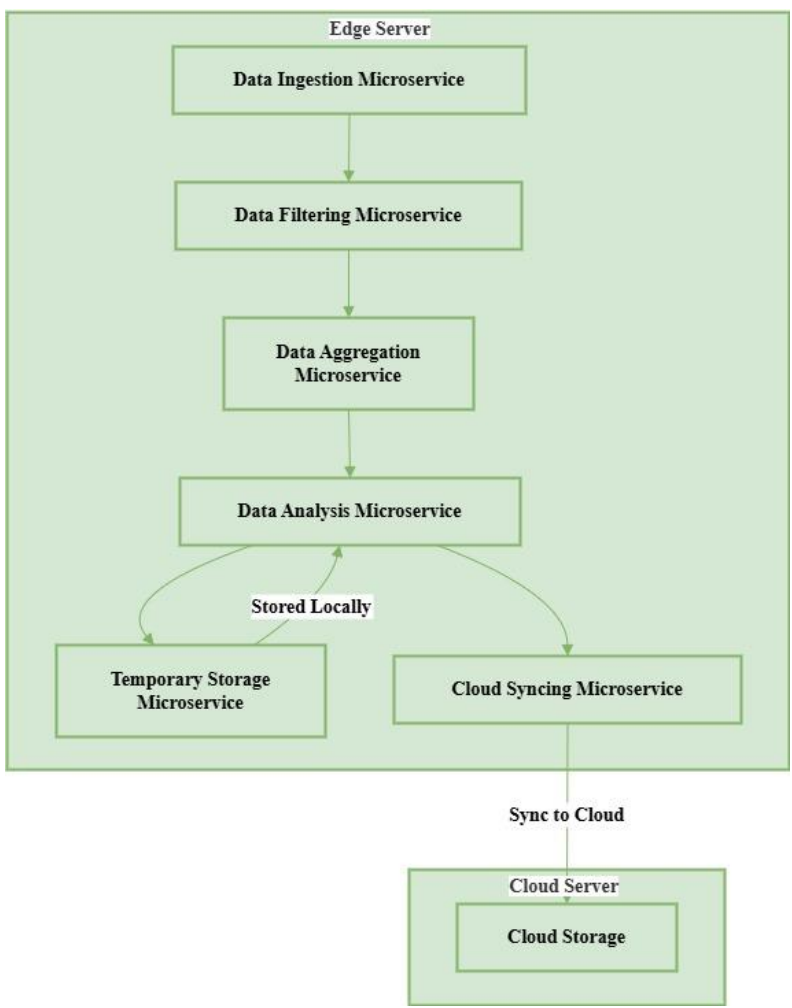

**Fig.4.** the diagram of the deployment of each microservice and their interactions within the system

**Microservices Deployment Flow Explained:**

- **The Data Ingestion Microservice (A)**: is responsible for receiving data from the edge devices and sending it to the filtering service.
- **The Data Filtering Microservice (B):** Entertainment Services It works on unprocessed information by eliminating unnecessary and superfluous details and transmitting the important data to the integrating microservice.
- **The Data Aggregation Microservice (C):** Merges the useful information into a short dataset and delivers it to the service for analysis.
- **The Data Analysis Microservice (D):** Evaluates the data brought together in aggregate for real time findings. It makes use of a storage medium (E) for holding data temporarily and makes use of cloud storage mechanism (F) for data that is meant for cloud preservation.
- **Temporary Storage Microservice (E):** Processes data and keeps it on standby while the system gets reoriented to the last accessed pieces of information.
- **Cloud Syncing Microservice (F):** Transfers the information to be stored in cloud server (G) for purposes of storage and other analytical activities.

This flow in figure 4 is illustrative of a microservices architecture which is modular and elastic in nature; every service performs a clearly defined task and all the tasks are well chained together in a manner suitable for real-time data processing at the edge and its efficient storage and analysis on the cloud.

### 3.5. Edge Computing Integration

The implementation of edge computing into airline ticketing systems involves placing a number of edge nodes within a geographic area to allow for the processing of important information nearer to the end users. This approach leads to less latency and enables real time interaction. The integration upholds the following major steps:

- **Segmentation of the Data Flows:**
  - In order to turn in reservation requests, clients queue their requests at the edge nodes instead of relying on some central system.
  - Accommodation is given for the operation of critical activities (such as the management of real time seat inventory and current pricing) at the edges and for non, current activities (such as the cloud storage of historical data for research) within the cloud.
- **Edge-Cloud Convergence:**
  - There is interworking of the edge nodes with the central cloud for the purposes of updates.
  - Large scale Big Data analytics processing is performed by centralized systems while edge systems are mainly concerned with time sensitive activities only.
- **Load Balancing and Dynamic Scaling:**
  - The edge architecture allows for changes in size especially during peak demand seasons.
  - The purpose of load balancing is to divide the loads among the edge servers.

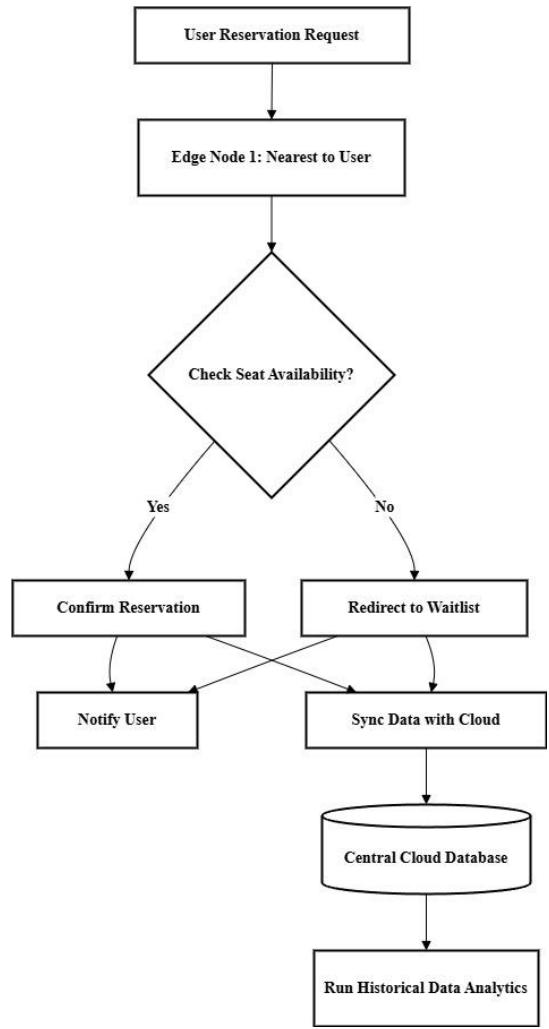

**Fig.5.** the Edge Computing Integration

**Description of diagram**

1. **Request Fulfillment of the Users**

To minimize the lag, user requests are always directed to the closest edge node.

2. **Edge node smart service processing**

Important functions such as checking the availability of seats and reserving them are completed at the edge.

3. **Cloud and Edge Device Interoperability**

The last part of data (like completed orders) is uploaded to the cloud database in order to maintain consistency across regions.

This approach in figure 5 improves system responsiveness, reduces latency, and balances the computational load between edge and cloud systems.

4. **Role of the Central Cloud**

The central cloud takes care of all the non-operational activities that are needed at the facility for example data analytics over large pots of data and forecasting.

## 4. Implementation and Tools

### 4.1. Technology Stack

The technology stack that would be required for an edge-aware microservices architecture in figure 6 and the airline reservation systems would be:

- **Containerization:**
    - Tools: Docker, Kubernetes.
    - Microservices, which are small, self-contained applications virtually built, are packaged and deployed using isolating containers which also allow extending and shrinking resources spread across an edge and a cloud.
- **Edge Computing Platforms and Tools:**
    - Tools: Google Anthos, Azure IoT Edge, and AWS Greengrass.
    - These platforms as well offer edge compute capabilities for users to process their data nearer to them in order to eliminate or reduce latency.
- **Messaging and Data Streaming:**
    - Tools: Apache Kafka, MQTT.
    - Messaging and streaming systems support edge and cloud nodes by enabling real time communication and coordination.

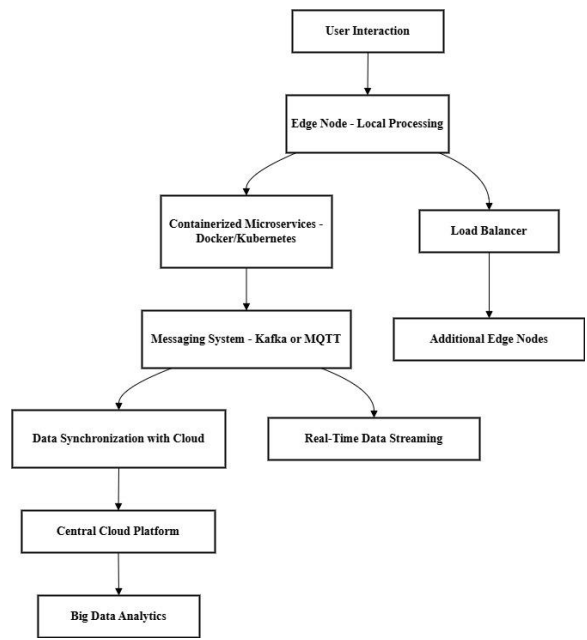

**Fig.6.** The technology stack for implementing an edge-enabled microservices architecture

**Explanation of Diagram**

- **User Interaction:**
    - Users make reservation requests or inquiries, which are sent to the closest edge node.
- **Edge Node Processing:**
    - Containers of microservices are located in edge nodes and provide real-time updates on seat availability and prices.
- **Messaging and Streaming:**

- The data are sent to the central cloud or other edge nodes by means of Kafka or MQTT for timely updating and analytics.
- **Cloud Integration:**
  - Less important data is transferred to the cloud for reasons of big data processing and storage.
- **Load Balancer :**
  - Provides a mechanism for load balancing by distributing incoming user requests evenly across edge nodes.

**Advantages of the Technology Stack**

- **Containerization:** Provides ability for quick, elastic deployment of microservices, and maintaining their homogenous environments.
- **Edge Computing:** Provides faster response times and improved experience for users.
- **Messaging Middleware:** Provides assured and timely services in communication in large scale distributed systems.

### 4.2. Integration Strategies

Incorporating edge appliances with cloud infrastructures and modernizing their environment to microservices comes to a careful consideration and a step-by-step procedure. Some strategies are listed below.

### 1. Associating Edge Devices and Cloud Backends

Strategy: Implement data protocols for the edges and provide connections to the cloud.

**Steps:**

- Create Systems and APIs for them to be synced in real time.
- Employ edges where data is regarded as processed and most importantly reduce cloud to avoid time waste.
- Conduct a data check at intervals, and dump data into the cloud to maintain data alignment.

### 2. Upgrading from Traditional Applications to Microservices

Strategy: Appetite for removing conventional systems a little at a time and replacing it with deploy –able microservices.

**Steps:**

- Don't begin with the critical modules but as the project is gradual, start with microservice refactoring of non-critical modules.
- Services should be deployed independently through containerization with tools such as docker.
- Use kubernetes an orchestration tool to take care of deployment and scaling.

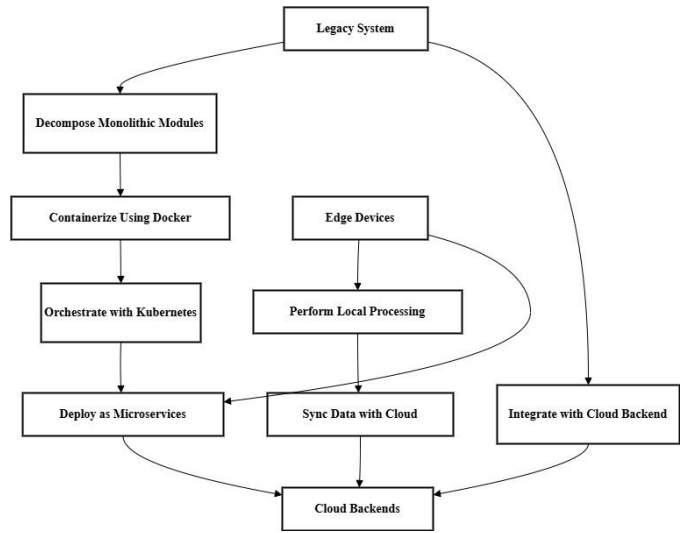

**Fig.7.** migrating legacy systems to microservices and integrating edge devices with cloud backends.

This is a flowchart in figure 7 depicting linear parallel processes, vertical processes are migrating current legacy systems into microservices and assimilating edge devices to the cloud back ends.

### 4.3. Development Workflow

The DevOps perspective is geared towards, and more importantly, automating the process of software delivery and the health of the system through CI/CD pipelines, and in addition, monitoring and observability concepts. this ensures that microservices are provisioned very often, and very little time is lost with provisioned resources as system visibility is at no point compromised.

**1. CI/CD Pipelines for Deployments that Need Regular Updates**

- Strategic imperatives: The process to prepare, test, and deliver or otherwise deploy a software product is articulated in a way it supports frequent software versions.

**Steps**

- Build: Application source code is written, and containerized applications (docker) built.
- Test: Unit, integration, and end to end tests - all performed automatically.
- Deploy: Deployment and management of applications done through Kubernetes.
- Watch: Monitor the deployment constantly and initiate a rollback in case of need.

**2. Monitoring and Observability Tools**

- Strategic imperative: Development of systems for health checks, system performance analysis, and general system production support.

**Systems:**

- Prometheus - for metrics collection.
- Grafana - for dashboards & visual representation.
- ELK - For logs and data searching, use ELK (Elasticsearch, Logstash, and Kibana).
- Jaeger / Zipkin - for traceability across distributed systems.

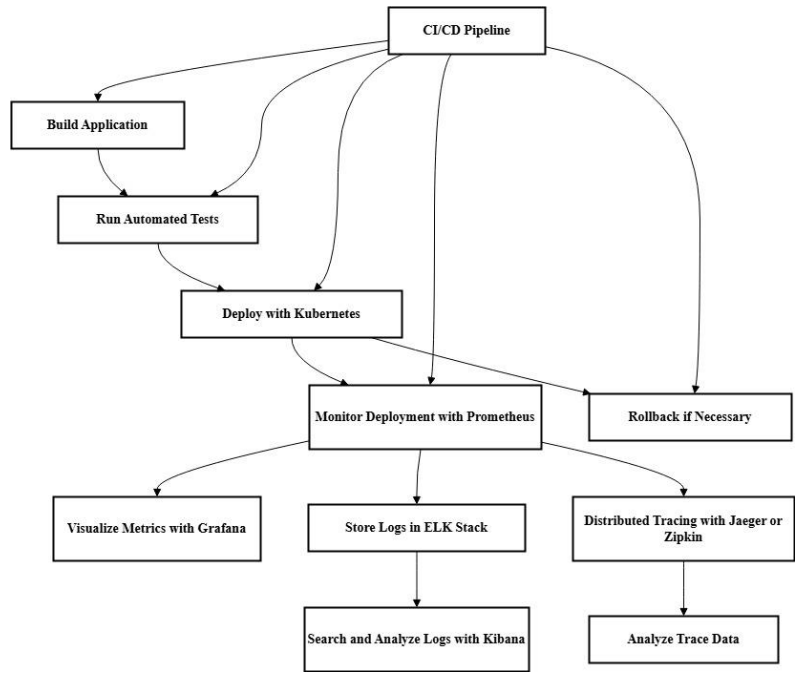

**Fig.8.** the CI/CD pipeline for continuous delivery and the observability tools

**Description**

**CI/CD Pipeline**

- Build: Code is built into deployable units (e.g., Docker images).
- Test: Automated tests (unit, integration) help to ascertain the quality of the code.
- Deploy: Deployments across environments (staging/production) are automated by Kubernetes.
- Monitor: Continuous Monitoring is established with Prometheus and alerting.
- Rollback: Kubernetes deployment rollback mechanism ensures that the user is taken back to a previously stable version automatically if a deployment fails.

**Monitoring and Observability:**

- Metrics are gathered and system status check is performed with Prometheus.
- Metrics with their interactive dashboards are presented by Grafana.
- To assist in issue resolution and system performance management, the ELK Stack is used for log monitoring.
- Request tracking and service performance watching is done using Jaeger/Zipkin by monitoring all requests and responses between services.

The following diagram in figure 8 illustrates the continuous delivery CI/CD pipeline and the corresponding observability tools for system health checks.

### 4.4. Evaluation Metrics

One can consider different metrics to assess an edge-enabled microservices framework. Latency refers to the time, which in this case is the period between the generation of information and the process of that information, with lower latency supporting better real-time interaction [37]. Data Throughput provides a measure of how much data is processed within a given time period and represents the ability of the system to sustain significantly high data volumes with no interruption (Xu, Liu, & Jiang, 2018). System Response Time encompasses both the latencies of the system and the speed of its data processing thus rendering an overall rating of the system which is important in applications that require instant response [36]. Finally, User Satisfaction includes the evaluation of the performance of the system from the eyes of the end-user which in particular are satisfaction scores and completion rates in order to ensure that the framework is able to serve its purpose [33].

## 5. Challenges and Solutions

### 4.5. Technical Challenges

Network Limitations and Latency Problems

**Challenges**

In situations of heavy usage of various connections, offering rapid services with possible transmission between edge nodes to the cloud can, at times, render results slower than intended. Usage in some geographies with unstable/fixed bandwidth connections makes interactivity of the system difficult.

**How to solve the problem**

- Content delivery networks (CDNs) should be used to store the most accessed data closer to the users.
- Similar to the above, introduce adaptive bitrate streaming or other data compression methods.
- Use private leased lines, or use 5G connectivity for reduced latency and reliable connection.
- Adopt load balancing strategies that are predictable in nature to decrease congestion levels within the network, particularly at peak activity periods.

Managing the Consistency of Distributed Data

**Challenges**

- Updating different edge nodes and the central cloud backend is highly sensitive due to possible updates and conflicting situations, which makes it difficult to keep data consistency.
- Using eventual consistency models presents the risk of temporary inaccuracies of key airline booking information (e.g. where seats are available).

**Solution**

- Combine event sourcing with CRDTs to achieve consistency over distributed computing systems.
- Implement strong consistency requirements especially for critical operations like reservation and cancelation transactions to ensure consistency at all time.
- Design conflict resolution solutions based on versioning and reconciliation in distributed databases.
- Implement data partitioning mechanisms to lessen the possibility of conflicts by allocating certain geographic regions or datasets to certain edge nodes only.

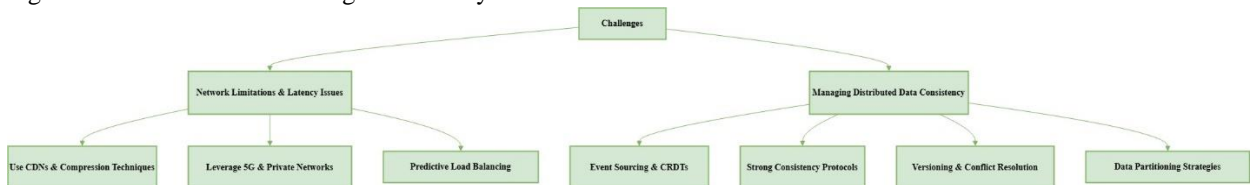

**Fig.9.** diagram of the primary technical challenges and actionable solutions

This part in figure 9 provides an overview of the main technical issues and the workable solutions available as way forward resolutions in order to enhance and boost the resilience as well as the system's overall performance.

### 4.6. Cost and Resource Challenges

#### 4.6.1. Edge Infrastructure Expenditure

- **Challenges**

- Infrastructure at the edges of the system inclusive of edge devices and storage and compute nodes is expensive both in terms of its installation and upkeep, especially in areas that are not well endowed with connectivity.
- Therefore, dedicated funds towards purchasing hardware and software platforms as well as running operational activities puts a significant burden on the available finances.
- **Proffered Solution**
  - Make use of cloud-edge services such as AWS Greengrass, Microsoft's Azure IoT Edge or Google Anthos which can be availed in a prepaid manner, thereby decreasing capital outlay.
  - Eliminate the need for building new edge infrastructure by exploiting the existing network assets in partnership with telecom operators.
  - Start with high-density population areas to enjoy the benefits and costs ratios.
  - Consider container-based approaches (for example: Docker, Kubernetes,...) at the edge to cut down on resources and costs.

### 4.6.2. Optimization of Resource Allocation

- **Challenge:**
  - Maintaining the right number of compute/storage resources at the edge nodes and in the cloud can be a quest.
  - While under-utilized resources can be an avenue for losing money, excess resources can lead to underperformance.
- **Solution:**
  - Employ resources resizing based on real-time demand using dynamic resource allocation algorithms.
  - Employ machine learning models to foresee the peak usage periods and adjust the resources for provisioning.
  - Storage optimization measures can be applied by using two approaches to storage where crucial information is stored at the edges and non-critical information is stored in the cloud.
  - Monitor resource consumption patterns and analyze them regularly with specific tools such as Prometheus and Grafana in order to make accurate adjustments of the use of resources.

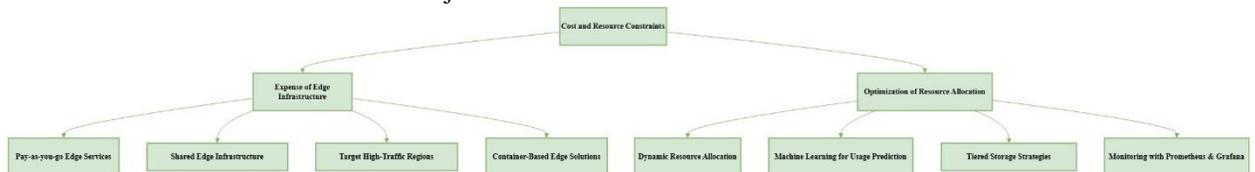

**Fig.10.** diagram of the cost and resource challenges and actionable solutions

This part in figure 10 provides an overview of the cost and resource issues and the workable solutions available as way forward resolutions in order to enhance and boost the resilience and the performance of the system under consideration.

## 4.7. Proposed Mitigations

In order to effectively deal with the issues that have been recognized, a variety of technical and strategic mitigation measures can be undertaken

### 4.7.1. Limitations of Network and Latency Issues

- **Mitigation Techniques**
  - **Edge Caching:** Store the data that is most frequently requested by users in the nodes at the periphery to avoid delays in recall.
  - **Content Delivery Network (CDN):** Content is replicated in several places to make it easily available to the end user.
  - **Load Balancing:** Employ predictive load balancing techniques in order to avoid node bottlenecks or overloading in one particular area of the network.
  - **Incorporate 5G:** Provision for high capacity locations in the network a combination of advanced connectivity solutions such as 5g.

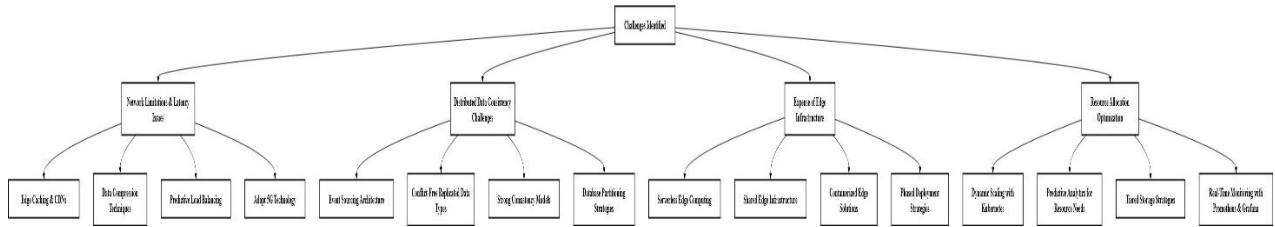

**Fig.11.** diagram of the cost and resource challenges and actionable solutions

- **Data Reduction:** Reduce the size of data packets for transmission purposes especially when sending data in a network that has limits on its capacity.

### 4.7.2. Ensuring consistency when handling data across distributed systems

- **Mitigation Techniques**
  - **Event Sourcing:** An architectural style which involves capturing changes as events and ensuring all nodes are consistent with them within an application.
  - **CRDTs:** Allow users to update such data even when they are not connected, and ensure that data updated while offline is automatically merged with no conflicts.
  - **Strict Consistency Level:** For certain types of data (reservations for example) requiring a lock or consensus prevents any updates from being executed until all nodes agree, locking the data for the entire process.
  - **Horizontal Data Partitioning:** Synchronization constraints can be alleviated if data is partitioned according to geo or functional zones.

### 4.7.3. Cost of Edge Infrastructure

- **Mitigation Techniques**
  - **Serverless Edge Computing:** Utilise edge services through serverless computing such as AWS Lambda@Edge to circumvent erecting new infrastructures.
  - **Rent instead of Build:** Work with cellular networks to install or use their edge devices and networks.
  - **Container Edge Solutions:** Offer service in virtualized container installed on the service provider's equipment cutting down on costs.
  - **Gradual Adoption:** Edge infrastructure and services will be adopted in stages bearing in mind the geographical areas with……..

### 4.7.4. Resource Allocation Enhancement

- **Mitigation Strategies**
  - **Dynamic Scaling:** Employ Kubernetes autoscaling capabilities to provision resources as the workloads fluctuate.
  - **Predictive Analytics:** Use AI/ML models to predict traffic and provision resources if necessary.
  - **Tiered Storage:** Use local storage at the edge for high-value data while keeping data that is less frequently accessed in the cloud.
  - **Real-Time Monitoring:** Application of Prometheus and Grafana to enhance the utilization of resources and eliminate wastages on the reallocation of resources.

**Description**

**Network Limitations and Latency Issues**: The focus of mitigations is to optimize data flows at the edge than processing in the cloud.

**Distributed Data Consistency:** Strategies focus on the management of conflicts and consistency of the nodes that are located within different geographical locations.

**Edge Infrastructural Costs:** Costs associated with edge infrastructure expenses are lowered by gradual enhancements, cloud services, and cooperation with other service providers.

**Resource Allocation Optimization:** Effective use of resources without wastage is achieved through the use of dynamic and predictive approaches.

This detailed mitigation plan covers in the figure 11 the issues that were highlighted in the previous parts of the document with practical measures that can be employed.

## 6. Conclusion

The addition of edge microservices into the airline reservation systems is a significant development in overcoming the shortcomings of standard centralized systems. The proposed33 framework is enhancing system responsiveness, decreasing latencies and bettering user experience with the edges for real-time processing and the microservices for the modular and the scalable systems.

This research shows key improvements in:

- **Real Time Data Processing:** Edge where necessary helps to manage operations which are critical such as checking for available seats and copying reservations.
- **System Scalability**: Allows the incorporation of dynamic scaling and other resource management practices to cope with heavy traffic especially at peak periods.
- **User Centric Architecture:** Improvement in speeds of response together with processing which is done nearer to the user improves the experience of users, thus very high levels of satisfaction and loyalty being reported.

The framework is also designed to meet the requirements of the changing airline market, incorporating advantages such as improved operational performance, cost efficiency and preparation for new technologies in the future. This method therefore provides a very effective basis for the modern airline system by solving the technical issues of networks, data distributions and resources management.

As the time passes and with the endless flow of advancement, the system as proposed in this paper can also be enhanced with the use of AI systems for advanced decision making, coupled with IoT and blockchain technologies for smart, efficient and safe systems. Moreover, this framework can also be modified to suit other sectors such as health care, management of transportation and even smart cities that makes it more dynamic and widely useful.

In summary, this research not only presents a solution to the existing problems concerning reservation systems in airlines. Nevertheless, it provides a flexible and scalable architecture which is capable of fast-tracking infrastructural development.

...

29. Xiang, H., Liu, H., Du, Y., Wang, X., Zhang, C., & Li, Y. (2021). No free lunch: Microservice practices reconsidered in industry. Proceedings of the ACM/IEEE 42nd International Conference on Software Engineering, 407-418.
30. Barua, B., & Kaiser, M. S. (2024). Leveraging Machine Learning for Real-Time Personalization and Recommendation in Airline Industry.
31. Chen, X., Mao, H., Zhang, H., & Li, D. (2019). Edge computing for internet of things applications: A case study on smart city deployment. Future Generation Computer Systems, 95, 375-389. https://doi.org/10.1016/j.future.2019.01.003
32. Barua, B., & Kaiser, M. S. (2024). Cloud-Enabled Microservices Architecture for Next-Generation Online Airlines Reservation Systems.
33. Li, J., Zhang, X., Huang, Z., & Chen, Q. (2020). Edge computing in industrial internet of things: Architecture, advances, and challenges. IEEE Communications Surveys & Tutorials, 22(4), 2743-2769. https://doi.org/10.1109/COMST.2020.3009562
34. Sarkar, S., Sinha, S., & Misra, S. (2021). Edge computing in smart cities: A comprehensive review. Journal of Network and Computer Applications, 186, 103063. https://doi.org/10.1016/j.jnca.2021.103063
35. Barua, B., & Kaiser, M. S. (2024). Leveraging Microservices Architecture for Dynamic Pricing in the Travel Industry: Algorithms, Scalability, and Impact on Revenue and Customer Satisfaction. arXiv preprint arXiv:2411.01636.
36. Satyanarayanan, M. (2017). The emergence of edge computing. IEEE Computer, 50(1), 30-39. https://doi.org/10.1109/MC.2017.9
37. Shi, W., & Dustdar, S. (2016). The promise of edge computing. IEEE Computer, 49(5), 78-81. https://doi.org/10.1109/MC.2016.145
38. Xu, Y., Liu, Z., & Jiang, Y. (2018). Real-time data processing for IoT-based edge computing. IEEE Access, 6, 24485-24496. https://doi.org/10.1109/ACCESS.2018.2826424
39. Barua, B., & Kaiser, M.S. (2024). Novel Architecture for Distributed Travel Data Integration and Service Provision Using Microservices.
40. Zhang, Y., Chen, M., Wu, Y., & Huang, Q. (2021). Leveraging edge computing for reducing latency in industrial IoT applications. IEEE Transactions on Industrial Informatics, 17(6), 4275-4283. https://doi.org/10.1109/TII.2020.3007597
41. Chaki, P. K., Barua, B., Sazal, M. M. H., & Anirban, S. (2020, May). PMM: A model for Bangla parts-of-speech tagging using sentence map. In International Conference on Information, Communication and Computing Technology (pp. 181-194). Singapore: Springer Singapore.
42. Li, J., Zhang, X., Huang, Z., & Chen, Q. (2020). Edge computing in healthcare: Enabling real-time patient monitoring and emergency response. IEEE Communications Surveys & Tutorials, 22(4), 2743-2769. https://doi.org/10.1109/COMST.2020.3009562
43. Barua, B., & Kaiser, M. S. (2024). A Next-Generation Approach to Airline Reservations: Integrating Cloud Microservices with AI and Blockchain for Enhanced Operational Performance. arXiv preprint arXiv:2411.06538
44. Chaki, P. K., Sazal, M. M. H., Barua, B., Hossain, M. S., & Mohammad, K. S. (2019, February). An approach of teachers' quality improvement by analyzing teaching evaluations data. In 2019 Second International Conference on Advanced Computational and Communication Paradigms (ICACCP) (pp. 1-5). IEEE.
45. Mphasis. (2018). Microservices for airlines industry whitepaper. Retrieved from Mphasis
46. Zeng, Y., Li, X., & Wang, Y. (2022). Online deployment algorithms for microservice systems with complex dependencies in cloud and edge computing. IEEE Transactions on Parallel and Distributed Systems, 33(5), 974-987. IEEE Xplore
22